\documentclass[aps,prl,twocolumn,noshowpacs,superscriptaddress,groupedaddress]{revtex4} 
\usepackage{graphicx}  
\usepackage{dcolumn}   
\usepackage{bm}        
\usepackage{amssymb}   

\usepackage[T2A]{fontenc}

\usepackage[utf8]{inputenc}

\usepackage{amsmath}

\usepackage{amsfonts}

\usepackage{textcomp}

\usepackage{bm}

\usepackage{color}

\usepackage{verbatim}

\usepackage{subfig}

\begin{document}

\title{Chiral vortical effect in pionic superfluid vs spin alignment of baryons}

\author{Oleg V. Teryaev}
\affiliation{JINR, 141980, Dubna, Russia \\ National Research Nuclear University (NRNU MEPhI)
31, Kashirskoye shosse, 115409 Moscow, Russia}
\author{Valentin I. Zakharov}
\affiliation{ITEP, B. Cheremushkinskaya 25, Moscow, 117218 Russia\\
School of Biomedicine, Far Eastern Federal University, Vladivostok, 690950, Russia\\
Moscow Inst. Phys. and Technol., Dolgoprudnyi, 141700 Russia}

\begin{abstract} 

We consider chiral fluids, with (nearly) massless fermionic constituents, in the confining phase.
Chiral vortical effect (CVE)  is the flow of axial current along the axis of rotation of the fluid
while the
spin alignment is a non-vanishing  correlation of polarizations of baryons
with the axis of rotation. 
As the theoretical  framework we use
 the model of pionic superfluidity induced by a non-vanishing isotopic chemical potential.
We note that
the average value of spin of virtual baryons reproduces the CVE. The role of defects,
or vortices is crucial. The model does not apply directly to the quark-gluon plasma
but might indicate existence of a mechanism to produce baryons with relatively large
polarization in heavy-ion collisions.

\end{abstract}
\maketitle

\subsection{Introduction}

Observation  of non-vanishing polarization of 
$\Lambda (\overline{\Lambda})$-hyperons in heavy ion collisions by the STAR Collaboration \cite{star}
is considered to be one of most remarkable experimental findings of recent years.
The common
interpretation nowadays is that spin of the hyperons is aligned with the
rotation of plasma  produced in peripheral ion-ion collisions.
Moreover, such a prediction was made prior to the experimental discovery,
(see, e.g. \cite{sorin,becattini} and references therein) and explored in some detail 
after the data appeared \cite{becattini2,sorin2}.

``Magnetization by rotation'' was introduced first about a century ago
\cite{century}. The origin of this phenomenon is readily understandable. Indeed,
with account of 
chemical potential $\mu_i$,
external magnetic field $\vec{B}$ and of rotation 
with angular velocity $\vec{{\omega}}$, the density operator $\hat{\rho}$
takes the form:
\begin{equation}\label{density}
\hat{\rho}~=~\frac{1}{Z}\exp\Big(-\hat{H}/T+\mu_i\hat{Q}_i/T+
\vec{\omega}\cdot\hat{\vec{J}}/T+\hat{\vec{\mu}}\cdot{\vec{B}}/T\Big)~,
\end{equation}
where $\hat{H}$ is the Hamiltonian, $T$ is temperature, $\hat{Q}_i$ 
are charges conjugated with $\mu_i$,
$\hat{\vec{J}}$ is the total angular momentum, $\hat{\vec{\mu}}$ is the magnetic moment. 
Thus, correlation of the total angular momentum $\vec{J}$ with $\vec{\omega}$ is quite obvious.
Spin $\hat{S}$ is a part of the total angular momentum. However, separation of the spin from the
orbital momentum is, generically, not a simple problem. Note, that the apperance of combination 
$L+S$ in the Dirac equation \cite{Hehl:1990nf} for the rotating frames  may be related \cite{Obukhov:2009qs} to the equivalence principle for Dirac fermions \cite{Kobzarev:1962wt,Kobsarev:1970qm}.

In case of  equilibrium of weakly interacting particles
evaluations with the density operator (\ref{density}) are quite straightforward,
for recent examples  and references see, e.g., \cite{becattini}.
An exception is, probably, the massless case
when one has to switch from the spin variables to chirality.
On the other hand, the limit of massless spin particles
is of special interest since there are new, chiral symmetries arising in this
limit. Alignment of chiralities of massless fermions with the magnetic field $\vec{B}$ 
results in the so called magnetic chiral effect (CME) \cite{vilenkin1}
which is a flow of electric current along the magnetic field. In case
of a single charged particle: 
\begin{equation}\label{cme} 
\vec{j}_{el}~=~\frac{e\mu_5}{2\pi^2}\vec{B}~,
\end{equation}
where $\mu_5$ is the chiral chemical potential, $\mu_5\equiv 1/2(\mu_R-\mu_L)$.
Another effect \cite{vilenkin2} is the flow of axial current along the axis of
rotation,  or $\vec{\omega}$:
\begin{equation}\label{cve}
\vec{j}_{A}~=~\frac{\mu^2+\mu^2_5}{2\pi^2}\vec{\omega}~~,
\end{equation}
where $\mu$ is the chemical potential associated with the
electric charge and we neglected temperature effects for simplicity.
 It is actually the  chiral (axial) vortical effect (\ref{cve})
that occupies our attention mostly.

In recent years, the chiral effects attracted a lot of attention,
see, e.g., the volume of review articles  \cite{volume}.
The progress was made
in direction of creating theory of systems much more complicated
 than free fermions and include, in particular, condensed-matter systems. 
Also, the chiral effects were recognized as macroscopic manifestations of
the anomalous triangle graphs. Instead of trying to compose a comprehensive
list of references let us just mention a breakthrough paper \cite{surowka}
where it was shown that Eq. (\ref{cve}) holds, in a slightly modified form,
for a generic liquid in hydrodynamic approximation. Namely,
Eq. (\ref{cve}) becomes
\begin{equation}\label{current}
j^{\mu}_A~=~\frac{\mu^2+\mu_5^2}{2\pi^2 }\omega^{\mu}~,~~~\omega^{\mu}=
\epsilon^{\mu\nu\rho\sigma}u_{\nu}\partial_{\rho}u_{\sigma}~,
\end{equation}
where $u_{\mu}$ is the 4-velocity of an element of liquid.

Despite of all the progress made,  it is not clear, how to apply 
theory to confining phases. 
And that is what is needed to appreciate
the results on hyperon polarization mentioned above. 
A crucial question is whether the spin orientation of quarks survives
the transition to confinement. 
A heuristic estimate
of polarization of the $\Lambda$-hyperon was
suggested in Ref. \cite {gorsky} in terms of vacuum
expectation values of chirality-changing (C-parity-odd) operators:
  \begin{equation}
<\sigma_z>~\simeq~\frac{<\overline{\Psi}\sigma_{xy}\gamma_5\Psi>}{<{\overline{\Psi}\Psi}>}
\end{equation}
 where $<\overline{\Psi}\Psi>$ is the quark condensate, 
and $<\overline{\Psi}\sigma_{xy}\gamma_5\Psi>$
is induced in the vacuum if external magnetic field or rotation is applied
(for details see \cite{gorsky})).

We are utilizing the model of pionic superfluidity induced by a chemical potential $\mu_3$
breaking isotopic invariance, see \cite{sonstephanov} and references therein.
It is not a realistic model for the quark-gluon plasma but this is a rare
example when the effects of confinement can be accounted for.
We will demonstrate that the average density of spins (polarizations) of baryons
 matches the density of axial current in the medium:
\begin{equation}\label{result}
<\sigma_z>_{baryons}~\approx~<\mu^2 \omega_z>~.
\end{equation}
 A crucial role in derivation of (\ref{result}) is played by defects,
or vortices. Formally, in case of superfluid
the vortices are infinitely thin.
  Baryons regularize the vortices at short distances. 
Note that vortices  have been considered in many papers.
Our treatment 
of vortices is close to that of papers \cite{sonzhitnitsky,metlitski,kirilin}.

In the next sections we give details of derivation of (\ref{result}). 

\subsection{Pionic superfluidity}

As is well known, at low energies, $E\ll \Lambda_{QCD}$ 
pions are representing the light, or physical degrees of freedom.
One can construct effective field theory describing the pion interactions which respects
the symmetries  of the original theory (QCD). Moreover, the effective action
is expandable
in derivatives, in the spirit of hydrodynamics. To make the problem tractable,
one introduces
non-vanishing chemical potentials, $\mu_V,\mu_A$ 
which violate isotopic symmetry and are
associated with the conserved charges $Q_V^3,Q_3^3$,
where the superscrit ``3'' stands for the third component of isotopic vector.
The temperature is kept, for simplicity, zero, $T=0$.
It was demonstrated, see
\cite{sonstephanov} and references therein, that using the general framework
of effective field theories,
one can show that the pionic liquid is superfluid. 
Let us recall the basic steps of the derivation \cite{sonstephanov}
emphasizing the points crucial for our considerations. 

 The effective Lagrangian is constructed in terms of unitary matrices $U$:
\begin{equation}\label{umatrix}
U~=~\exp (i{\lambda^a\pi^a}/{f_{\pi}})
\end{equation} 
where $\pi^a$ are massless Goldstone fields 
\footnote{Note that we keep pion mass vanishing. In fact, the effect
of small masses could be lso included, see \cite{sonstephanov,zamaklar}.}
associated with the spontaneous 
breaking of the chiral symmetry, $\lambda^a$ are Hermitian matrices.
The effective Lagrangian incorporating the chemical potential $\mu_V, \mu_A$ looks as
\cite{sonstephanov,zamaklar}:
\begin{equation}\label{lchiral}
L_{chiral}=~\frac{f_{\pi}^2}{4}\Big(D_{\mu}UD^{\mu}U^{\dagger}\Big)~,
\end{equation}
where the covariant derivatives are defined as:
\begin{eqnarray}
D_{\mu}U~=~\partial_{\mu}U
~-~i\delta_{\mu 0}\big(\hat{\mu}_LU-U\hat{\mu}_R\big)~~,
\end{eqnarray}
and we will consider only $\hat{\mu}_L, \hat{\mu}_R~
\equiv~\mu_L\sigma_3, \mu_R\sigma_3$, where $\hat{\mu}_{L,R}$ are matrices
and $\mu_{L,R}$ are numbers. 
The next step is to minimize the potential energy  $V_{chiral}$:
\begin{equation}
V_{chiral}~=~\frac{f_{\pi}^2}{8}\big((\mu_V^2-\mu_A^2)Tr (U\sigma_3U^{\dagger}\sigma_3)
-(\mu_V^2+\mu_A^2)Tr I\big)
\end{equation}
One 
 can readily see that in case of $\mu_V= 0, \mu_A \neq 0$ the minimum
is reached on matrices $U$ of the form:
\begin{equation}
U_{min}~=~I \cos\phi~+~\sigma_3 \sin\phi  ~,
\end{equation}
where $\phi$ is an angle and  independence of the energy on $\phi$ 
signals presence of a Goldstone boson. In other words, we have a trivial vacuum in this
case:
\begin{equation}\label{case1}
U_{min}~=~I~,~~~if ~\mu_A~\neq~0, ~\mu_V~=~0 ~~.
\end{equation}
Similarly, 
\begin{equation}\label{case2}
U_{min}~=~\frac{1}{\sqrt{2}}(\sigma_1+
i\sigma_2)~,if~\mu_V~\neq~0,~\mu_A~=~0~,
\end{equation}
where $\sigma_{1,2}$ are Pauli matrices.

Although the cases (\ref{case1}) and (\ref{case2}) look different they can be reduced
to each other by a change in notations, or by global chiral rotations of matrices
$U_{min}$. The point is explained in detail in Ref. \cite{zamaklar},
see in particular Fig. 2 of this paper. In particular, the $U(1)$ subgroup which
rotates $\pi_1$ and $\pi_2$ into each other is replaced by the $U(1)$ subgroup
that rotates scalar $\sigma$ and $\pi_3$ into each other. We will concentrate on
the case $\mu_A\neq 0, \mu_V=0$.

So far we tacitly assumed that the kinetic terms vanish identically, since we considered
minimum of energy. However the covariant derivative $(D_0U)$ does not vanish if the matrix
$U$ does not depend on time. Instead, we
should rather look for a solution of equations of motion. To be more specific, turn to the
case (\ref{case1}) and consider
the matrix $U_{solution}$ explicitly depending on time $t$ \footnote{We follow here
\cite{avdoshkin}.}:
\begin{equation}\label{cartan}
U_{solution}~=~\exp \big(i\sigma_3\mu_A\cdot t\big)~~,
\end{equation}
where $\mu_A$ is now a number.
  Clearly,
\begin{equation}
D_0U_{solution}~=~0~,
\end{equation}
 It is also important that the matrix (\ref{cartan}) is constructed entirely
on the Cartan subgroup, i.e. matrices $I,\sigma_3$. 

Note that the proportionality
to time of the phase of the condensate is a well known signature of superfluidity,
see, e.g., \cite{nicolis}. Indeed, the criterion of superfluidity
is a specific form of the correlator of $T_{0i}$
components of the energy-momentm tensor:
\begin{equation}\label{correlatorr}
<T_{0i}, T_{0k}>_{superfluidity}~=~(const)\frac{q_iq_k}{q_i^2}~~.
\end{equation}
The solution (\ref{cartan}) implies
that the $\pi_0$-field looks as, 
\begin{equation}\label{fieldpi}
\frac{\pi_0}{f_{\pi}}=\mu\cdot t+\varphi(x_i),
\end{equation}
where the notation ``$\pi_0$'' is chosen to
make connection with the matrices  $U$ (\ref{umatrix}) explicit,
and the field $\varphi(x_i)$ satisfies equation $\Delta\varphi~=~0$.
It is then obvious that the criterion (\ref{correlatorr}) is fulfilled.
Indeed the $T_{0i}$ component for a scalar field contains term
$T_{0i}~\sim~\partial_t\phi\partial_i\phi$ and in our case
$T_{0i}~\sim~\mu\partial_i\varphi$
where $\varphi$ is a 3d  massless field.

 In conclusion of this section let us
emphasize that we consider the case of a ``small" chemical potential. Namely, introduction of a small
chemical potential does not change
the absolute value of the vacuum condensate but only rotates it. 
In other words, one actually assumes that
the potential energy contains a term like
\begin{equation}
V~\sim~M\Big(UU^{\dagger}~-~I\Big)~,
\end{equation} 
with $M\gg \mu_{L,R}$.
The density $n$ of particles in the condensate is calculable by taking derivative from 
the energy with respect to the chemical potental;
\begin{equation}\label{density}
n_{V,A}~\sim~\frac{\partial}{\partial \mu_{V,A}}V_{chiral}~=~2f_{\pi}^2\mu_{V,A}~.
\end{equation}
That is, the density disappears with $\mu_{V,A}\to 0$.
 
\subsection{Vortices}

Vortices in superfluid represent, probably,
the earliest example of ``defects'', see, e.g. ,
textbooks \cite{landau,pines}. General formalism 
was adapted to the case of relativistic superfluidity in many papers
as well,
see, e.g., \cite{nicolis,zahed,kalaydzhyan,kirilin}.  
  
Let us first recollect most general properties of the vortices \cite{landau}.
The velocity $v_i$ of the superfluid component is related to the gradient of the field 
$\varphi$ entering Eq. (\ref{fieldpi}):
\begin{equation}\label{gradient}
\partial_i\varphi~=~\mu v_i~~.
\end{equation} 
Note that mostly we are using non-relativistic notations.
Fully relativistic equations are always possible to introduce
in terms of $u_{\mu}$.

Because of  Eq. (\ref{gradient}) , naively, rotation of the superfluid is not allowed.
However the angular momentum is still transferred to the liquid through vortices
which are singular on the ($z$) axis. Namely, near the  axis:
\begin{equation}\label{n}
\pi_0/f_{\pi}~=~\mu t+n\theta
\end{equation}
where $\theta$ is the polar angle in plane 
orthogonal to the axis, $0\le \theta \le 2\pi$ and $n$ is integer.
Clearly, the angle $\theta$ is not defined on the axis. 
This is reflected in a singularity:
\begin{equation}\label{singularity}
(\partial_x\partial_y-\partial_y\partial_x)\theta~=~2\pi \delta(x,y)~,
\end{equation}
which is responsible for the transfer of rotation to the superfluid.

Let us also remind the reader, how the current 
 (\ref{current}) is related to the chiral anomaly. The point is that
 introduction of the chemical potential $\mu$ assumes replacing the original 
 Hamiltonian  $\hat{H}_0$ by $\hat{H}_0-\mu Q$, where $Q$ is the
 charge associated with the potential $\mu$. In case we are considering
\begin{equation}\label{delta}
\delta \hat{H}~=~-\mu\int d^3x\overline{\Psi} \gamma_0\Psi~-~\mu_5\int d^3x 
\overline{\Psi}\gamma_5\gamma_0\Psi~,
\end{equation} 
where $\Psi$ stands  for a generic massless spin 1/2 field.

Generalization, of (\ref{delta}) to the case of liquid is achieved, as usual,
by using the 4-vector $u_{\mu}$. In terms 
of the density of the Lagrangian we then have:
\begin{equation}\label{noval}
\delta L~=~\mu u_{\alpha}\overline{\Psi}\gamma^{\alpha}\Psi+
\mu_5u_{\alpha}\overline{\Psi}\gamma_5\gamma^{\alpha}\Psi~.
\end{equation} 
 Note the similarity of the novel, specific for thermodynamics
 interaction (\ref{noval}) with the ordinary electromagnetic interaction,
 $\delta L_{el}=~eA_{\alpha}\Psi\gamma^{\alpha}\Psi$. This similarity
 implies that there is an extension of the standard chiral triangle anomaly
which can be 
 generated \cite{shevchenko} by the substitution:
 \begin{equation}\label{substitution}
 eA_{\alpha}~\to ~eA_{\alpha}+\mu u_{\alpha}+\mu_5 u_{\alpha}.
 \end{equation}
 Keeping, for simplicity, $\mu=0, \mu_5\neq 0$ we come to the anomalous
 current $j^{\alpha}_5~=~(\mu_5^2/2\pi^2)\omega^{\alpha}$, where
 $\omega^{\alpha}$ is defined  in
 the Eq. (\ref{current}).
 
 After these preliminary remarks, we are in position to evaluate the 
 anomalous current
 (\ref{current}) in our case of pionic superfluidity.
This is in three simple steps.   First, 
the generaliztion of the triangle graph  
produces the current:
\begin{equation}
j^{\mu}_5~=~\frac{1}{4\pi^2 f_{\pi}^2}\epsilon^{\mu\nu\rho\sigma}(\partial_{\nu}\pi^0)
(\partial_{\rho}\partial_{\sigma}\pi^0)
\end{equation}
The $\pi_0$  field near the axis of a vortex looks 
as given in Eq.(\ref{n}). As a result the current induced by the given vortex is:
\begin{equation}\label{linear}
j^{i}_5~=~\frac{\mu n}{2\pi} ~ \delta(x_{i+1},x_{i+2})~~~~(i=1,2,3)
\end{equation}
At first sight, this result is in variance with Eq. (\ref{cve})
since the current (\ref{linear}) is linear in the chemical potential $\mu$ while
the current (\ref{cve}) is quadratic in $\mu$.
However, to check the dependence on $\mu$ we have to express
the current (\ref{linear}) in terms of the same angular velocity $\vec{\omega}$
as used in Eq. (\ref{cve})
and we will see in a moment that this removes 
the apparent discrepancy between (\ref{linear}) and (\ref{cve}).

 The fact that $n$ is integer manifests quantization
of the angular momentum carried by the vortex. 
Indeed, from uniqueness of the phase of the
wave function one concludes 
\begin{equation}
\oint \partial_i\varphi dx_i~=~2\pi n
\end{equation}
and the corresponding value of $n$
(assuming again that vorticity is directed along the 3rd axis) 
is given by 
\begin{equation}\label{omega}
n=~\frac{1}{2\pi}\oint dx_i\partial_I\varphi~=~\frac{\mu}{\pi}\int d^2x\omega^3~,
\end{equation} 
where we used $curl \vec{v}~=~2\vec{\omega}$.
Substituting (\ref{omega}) into (\ref{linear})
and taking into account that vorticity is non-zero only on the vortices
we find out that Eq. (\ref{linear}) matches (\ref{cve}) exactly.

Note that the different sign of topological number $n$ 
for the antiparticles, when sign of chemical potential changes,
leads to the same sign of the axial current, in accordance 
to its positive C-parity.
This is compatible to the same signs of polarization of 
$\Lambda$ and $\bar \Lambda$ hyperons observed experimentally \cite{star}.

Note that rewriting the current (\ref{linear}) in the form similar to (\ref{cve})
is reasonable in case of a particular physical set up. Namely, 
it is assumed that a bucket
with superfluid is rotated with angular velocity $\vec{\omega}$
which is kept constant. 
As is mentioned above,
naively, the rotation is not transferred to the
fluid because of the condition $\vec{v}=\vec{\nabla} \varphi$.
However, this naive picture does not hold because of the 
singularity (\ref{singularity}) and there is a net of vortices in
the fluid which carry the same angular momentum
as if the liquid were rotated as a whole. Each particular vortex has
$n=1$ since such a configuration minimized the energy
 \cite{landau,pines}.
The density of vortices per unit surface is then fixed by the condition
that the whole of the angular momentum is carried by the vortices
and can be determined, say, from Eq. (\ref{omega}).

\subsection{Chiral effects in superfluid as radiative corrections}

As is mentioned above, chiral effects in superfluid have been considered
in many papers. For our purposes and interpretations it is 
instructive to view the  chiral effects 
as radiative correctins to the Born approximation
which is nothing else but the superfluidity itself.
This viewpoint was emphasized recently in \cite{avdoshkin}.

Superfluidity induced by the chemical potential
$\mu_A$ (see \cite{sonstephanov} and a brief summary above)
can be described as existence of the current $(j_{\alpha}^5)_s$:
\begin{equation}\label{superfluid}
(j_{\alpha}^5)_{s}~=~2f_{\pi}^2\mu_A (u_{\alpha})_{s}~\equiv~
n_A(u_{\alpha})_{s},
\end{equation}
where we utilize Eqs (\ref{density}) and, for the sake of definiteness, consider
the case $\mu_A\neq 0, \mu_V=0$. Moreover, $(u_{\alpha})_s$
is the 4-velocity of the superfluid component. Note that
the current (\ref{superfluid}) looks as a standard hydrodynamic current,
$j_{\alpha}=n\cdot u_{\alpha}$.

Then the chiral magnetic effect is readily recongnizable as a one-loop 
quantum correction to (\ref{superfluid}) induced by
electromagnetic interaction:
\begin{equation}\label{radiative}
j_{\alpha}^{el}~\sim~f_{\pi}^2\mu_A\cdot \frac{e}{4\pi^2}\frac{B_{\alpha}}{f_{\pi}^2}~~,
\end{equation}
where $B_{\alpha}$ is the magnetic field, 
$B_{\alpha}=\epsilon_{\alpha\beta\gamma\delta}u^{\beta}F^{\gamma\delta}$ and
the factor $(4\pi^2)^{-1}$ represents a ``typical'' numerical factor associated
with a loop graph. On the technical side, the precise form of Eq. (\ref{radiative})
can be obtained by using the Goldstone-Wilczek current
\cite{goldstone} (see also \cite{harvey}).
  
It is worth emphasizing  that the radiative correction (\ref{radiative})
comes from {\it short} distances. And the current (\ref{radiative})
is to be considered as a polynomial. Indeed, the light particles are
represented entirely
by pions. The Born-approximation current (\ref{superfluid})
is due to the flow of pions. On the other hand, the current (\ref{radiative})
cannot be supported by pions alone. Indeed, electromagnetic current of pions
is a component of  an isotopic vector. While to get a non-vanishing
contribution (\ref{radiative}) one needs an isoscalar component
of the electromaagnetic current as well. This isoscalar component
is manifested in the electromagnetic current associated with baryons,
like proton and neutron. Within the framework considered $m_{\pi}/m_B \to 0$
and (\ref{radiative}) is a polynomial.

Turn now to the chiral vortical effect. In this case we
consider rotating superfluid.  
Then, there is a radiative correction to (\ref{superfluid})
associated with the effective interaction $\mu_A u_{\alpha}$,
see discussion after Eq. (\ref{substitution}).
The radiative correction induces an axial-vector current
with density:
\begin{equation}\label{radiative2}
\delta j^5_{i}~\sim f_{\pi}^2\mu_A \frac{\mu_A\vec{\omega_i}}{4\pi^2 f^2_{\pi}}~,
\end{equation}
see Eqs (\ref{linear}), (\ref{omega}).
Alternatively, and as discussed above the radiative correction 
on average generates density of angular momentum $\vec{\mathcal{M}}$:
\begin{equation}\label{radiative1}
\vec{\mathcal{M}}~\sim~ \delta \vec{j}_5.
\end{equation}

Now, we come to a central point. The same as in the preceding example, the
correction (\ref{radiative1}) is associated with short distances.  
Because of this,
the  current (\ref{radiative2}) can be identified as the
current of heavy particles, or baryons. Moreover,
the spatial component of the axial current reduces to the
density of spin of baryons, $\delta j^5_i~\sim~<\sigma_i>$.
This conclusion fits nicely (\ref{radiative1}).
In other words, the core of the vortices is built on spins of heavy particles.
For a similar construction see \cite{sonzhitnitsky,metlitski}.
 
Note that vortices in the pionic superfluid
were considered, in particular, in Refs. \cite{kirilin,sadofyev}.
It was assumed that there are massless quark modes propagating along the
axes of the vortices. In other words, the confining 
vacuum is destroyed and perturbative vacuum is seen on the rotation axes.
However, as is emphasized in the discussion concluding section 2,
destroying the pion condensate does not mean that we reach
deconfining, or perturbative vacuum.

For simplicity, we limited ourselves to the pionic fluid with the respective isotopic chemical potential.
Because of the relation of $\Lambda$ polarization to that of strange quarks \cite{sorin}, the octet contribution to chemical potential may also play a role, although it is not obvious how realistic can be such  more complicated  model anyway. 

In conclusion of this section, let us discuss fixation of the overall
coefficients 
in expressions for the chiral effects.
The point is that the Goldstone-Wilzcek currents,
which are central to evaluate the chiral effects,
are finite. It is expressed in terms of Goldstone particles
which interact with heavy particles. Superficially, 
 the overall coefficient could
depend on how many species
of heavy particles, or baryons are kept.
However, the overall coefficients are determined from the matching 
to the   triangle anomalies which depend on quantum numbers 
of quarks in the underlying field theory.
In particular, the vortex-related baryonic current
is given by (see, e.g., \cite{kalaydzhyan}):
\begin{equation}\label{baryonic}
(j_B^5)_{\alpha}~=~\frac{N_c}{36\pi^2f_{\pi}^2}\epsilon_{\alpha\beta\gamma\delta}
(\partial^{\beta}\pi^0)(\partial^{\gamma}\partial^{\delta}\pi^0)~,
\end{equation}  
where $N_c$ is the number of colors. 

\subsection{Discussion}

Recent measurement of $\Lambda (\overline{\Lambda})$ polarization
motivate further theoretical studies of the chiral vortical effect
in the confining phase and of the role played in it by heavy particles.
We considered the model of pionic superfluidity
induced by isotopic-symmetry violating chemical potentials, see
\cite{sonstephanov} and references therein. This is not a realistic model
of the quark-gluon plasma but it allows for a remarkably simple and
convincing description of a hadronic medium.

Apriori, the impression is that this model is not suited at all to address
the issue of evluation of
the polarization of baryons since the model {\it per se} does not mention
heavy particles.  However, a more detailed consideration does reveal
that heavy particles do play a role as an ultraviolet cut off
in radiative corrections. The radiative corrections to superfluidity,
in turn, describe the chiral magnetic and vortical effects.  
In hydrodynamic terms, the radiative corrections correspond to
higher orders in derivatives.
Introduction of the ultraviolet cut off is crucial to regularize
the singularity on the axes of vortices. 

Thus, one starts with a model which includes only light, Goldstone particles
and comes to the conclusion that the model is inconsistent without
invoking heavy particles, or baryons as an ultraviolet cut off.
Moreover, the whole of the chiral vortical effect
is associated with the spins of heavy particles.
In this sense and qualitatively, the polarization of
baryons turns to be large.
 
Thus, the vortex in superfluid looks as follows,
see also \cite{sonzhitnitsky}. On the periphery of the vortex
there are light degrees of freedom, circling the core. The core is built on the polarization 
of heavy particles. This picture is a kind of complementary to the vortex
considered by Callan and Harvey \cite{harvey}. In the latter case the core is circled by
spins of heavy particles, with electric current flowing towards the core. The core itself
is occupied by light (massless) degrees of freedom. Finally, in case of magnetic vortices
magnetic moment is curling the core on the periphery and look along the core in the middle.

There are many reservations to applying the picture emerging to the realistic plasma.
First of all, the heavy particles in case of pionic superfuidity are rather virtual, not real.
Then, it is also important that the baryonic current (\ref{baryonic}) is an isotopic scalar
and its consideration is somewhat ambiguous because 
of the gluonic anomaly. Nevertheless, one might hope that
vortices in pionic superfluid medium provide us with an example
of mechanism of transformation of rotation of plasma to polarization of
baryons. Also, vortices might be relevant to the quark-quark gluon plasma
since the viscosity of the plasma is known to be low.

\subsection{Acknowledgments}

We are thankful to V.P.~Kirilin, A.V.~Sadofyev, J.~Sonnenschein and A.S.~Sorin for interesting discussions.
The work was supported by Russian Science
Foundation Grant No 16-12-10059.

\end{document}